\documentstyle[11pt]{article}
\begin{document}
\begin{titlepage}
\hfill{UQMATH-93-05}
\hfill{hep-th/9307007}
\vskip.3in
\begin{center}
{\huge Quantum Affine Algebra and Universal $R$-Matrix with Spectral Parameter}
\vskip.3in
{\Large Yao-Zhong Zhang} and {\Large Mark D. Gould}
\vskip.3in
{\large Department of Mathematics, University of Queensland, Brisbane,
Qld 4072, Australia}
\end{center}
\vskip.6in
\begin{center}
{\bf Abstract:}
\end{center}
Using the previous obtained universal $R$-matrix for the quantized
nontwisted affine Lie algebras $U_q(A_1^{(1)})$ and $U_q(A_2^{(1)})$,
we determine the explicitly spectral-dependent universal $R$-matrix for
the corresponding quantum Lie algebras $U_q(A_1)$ and $U_q(A_2)$.
As their applications, we reproduce the well-known
results in the fundamental representations and we also
derive an extreamly explicit formula of the
spectral-dependent $R$-matrix for the adjoint representation of
$U_q(A_2)$, the simplest non-trival case when the tensor product
decomposition of the representation with itself has finite multiplicity.

\vskip 6cm
\noindent {\bf Mathematics Subject Classifications (1991):} 81R10, 17B37, 16W30
\end{titlepage}

\noindent {\bf 1. Introduction:}
In this Letter, we present two main results: (i) the universal $R$-matrix
for quantum Lie algebras $U_q(A_1)$ and $U_q(A_2)$ with explicit spectral
parameter dependence; (ii) an extreamely
explicit formula for the spectral-dependent $R$-matrix in the adjoint
representation of $U_q(A_2)$, the simplest nontrival case when the
tensor product decomposition of the representation with itself has
finite multiplicity.

Since their invention, quantum algebras \cite{Drinfeld}\cite{Jimbo}
\cite{Reshetikhin} have been employed, as a main technique, to find spectral
parameter dependent solution to quantum Yang-Baxter equation (QYBE)
(that is spectral-dependent $R$-matrix). Two conventional approaches
to this effect in literatures may be viewed as the following two
seemingly different but essentially related
"Yang-Baxterization" processes: (i) one starts
from a quantum simple Lie algebra, affinizes it, thus giving rise to the Jimbo
equations, and then solves the Jimbo equations \cite{Jimbo}\cite{ZGB};
(ii) one starts with some braid group representation of a quantum simple Lie
algebra and then tries to introduce a spectral parameter in such a way that
the spectral-dependent QYBE is satisfied \cite{Jones}.
The "abelian Yang-Baxterization",
where the tensor product decomposition of representation with itself
is multiplicity-free, is extensively studied in literatures for lower
representations. For higher representations, the "fusion" procedure
is suggested which is not, however, easily workable. As to "non-abelian
Yang-Baxterization" in which the tensor product decomposition
of representation with itself has finite multiplicity, very
few attempt has currently been made, due to conceivable
complexities caused by the appearance of finite multiplicity. Therefore, it
seems desirable to develop more effective procedures to cope with the
situations.

In this Letter we will present a new way of obtaining spectral-dependent
$R$-matrix for quantum Lie algebras.
Our idea reverses, in some sense, the above process. Namely, we start from
the universal $R$-matrix of quantum affine Lie lagebra
$U_q({\cal G}^{(1)})$ and then apply it to
finite-dimensional loop representations $V(z)$ of $U_q({\cal G}^{(1)})$, which
is isomorphic to the ones $V\otimes {\bf C}(z,z^{-1})$ of the corresponding
quantum simple Lie algebra $U_q({\cal G})$.
In this way, a spectral parameter appears automatically and
we obtain the spectral-dependent solution to QYBE for the
latter. A remarkable advantage lying this approach is that both the case of
multiplicit-free and the case with finite mulitiplicity can be treated in a
unified fashion. As a matter of fact, we are able to get a spectral-dependent
{\em universal} form of $R$-matrix for $U_q(A_1)$ and $U_q(A_2)$, regardless of
representations and multiplicites considered. Applying to some
concrete representations,  we are able to reproduce the well-known
results in the fundamental representation and to obtain
an extremly explicit formula
for $R$-matrix of $U_q(A_2)$ for the adjoint representation.
\vskip.1in
\noindent {\bf 2.
Universal $R$-Matrix for $U_q(A_1^{(1)})$ and $U_q(A_2^{(1)})$:}
For self-contained, this
section is devoted to a brief and quick review of the
construction of the universal $R$-matrix for $U_q(A_1^{(1)})$ \cite{KT}
and for $U_q(A_2^{(1)})$ \cite{KT}\cite{ZG}.
Throughout the paper, we use the notations:
$({\rm ad}_qx_\alpha)x_\beta$=$[x_\alpha\,,\,x_\beta]$=$x_\alpha x_\beta
-q^{(\alpha,\beta)}x_\beta x_\alpha$\,,~~
$\theta(q^h)=q^{-h}$\,,~~$\theta(E_i)=F_i$\,,~~$\theta(F_i)=E_i$
\,,~~$\theta(q)=q^{-1}$\,,~~
$(n)_q=(1-q^n)/(1-q)$\,,~~$[n]_q=(q^n-q^{-n})/(q-q^{-1})$\,,~~
$q_\alpha=q^{(\alpha,\alpha)}$\,,~~
${\rm exp}_q(x)=\sum_{n\geq 0}\,\frac{x^n}{(n)_q!}$\,,~~$(n)_q!=
  (n)_q(n-1)_q\,...\,(1)_q$.

We start with rank 2 case.
Fix the normal ordering in the positive root system $\Delta_+$ of $A_1^{(1)}$ :
$\alpha$,\, $\alpha+\delta$,\, $\cdots$,\, $\alpha+n\delta$,\, $\cdots$,\,
$\delta$,\, $2\delta$,\, $\cdots$,\,
$m\delta$,\, $\cdots$\,,\, $\cdots$\,,\, $(\delta-\alpha)+l\delta$,\, $\cdots$
\,, \, $\delta-\alpha$ \,,
where $\alpha\,,~\delta-\alpha$ are simple roots and $\delta$ is the
minimal positive imaginary root. One has \cite{KT}
the universal $R$-matrix for $U_q(A_1^{(1)})$:
\begin{eqnarray}
R&=&\left (\prod_{n\geq 0}\;{\rm exp}_{q_\alpha}((q-q^{-1})(E_{\alpha+n\delta}
  \otimes  F_{\alpha+n\delta}))\right )\nonumber\\
& &\cdot{\rm exp}\left ( \sum_{n>0}n[n]_{q_\alpha}^{-1}
  (q_\alpha-q_\alpha^{-1})(E_{n\delta}\otimes F_{n\delta})\right )\nonumber\\
& &\cdot\left (\prod_{n\geq 0}\;{\rm exp}_{q_\alpha}((q-q^{-1})
  (E_{(\delta-\alpha)+n\delta}\otimes
  F_{(\delta-\alpha)+n\delta}))\right )\cdot
  q^{\frac{1}{2}h_\alpha\otimes h_\alpha+c\otimes d+d\otimes c}\label{sl2R}
\end{eqnarray}
where $c=h_\alpha+h_{\delta-\alpha}$ and the Cartan-Weyl generators,
$E_\gamma\,,~~F_\gamma=\theta(E_\gamma)\,,~~\gamma\in\Delta_+$\,, are given by:
$\tilde{E_\delta}=[(\alpha,\alpha)]_q^{-1}[E_\alpha,\,E_{\delta-
\alpha}]_q$\,,~~
$E_{\alpha+n\delta}=(-1)^n\left ({\rm ad}\tilde{E_\delta}\right )^nE_\alpha$
\,,~~$E_{(\delta-\alpha)+n\delta}=\left ({\rm ad}\tilde{E_\delta}\right )^n
E_{\delta-\alpha}$\,,~~
$\tilde{E}_{n\delta}= [(\alpha,\alpha)]_q^{-1}[E_{\alpha+(n-1)\delta},
\,E_{\delta-\alpha}]_q$ and
\begin{equation}
\tilde{E}_{n\delta}=\sum_{p_1+2p_2+\cdots+np_n=n}
  \frac{\left ( q^{(\alpha,\alpha)}-q^{-(\alpha,\alpha)}\right )^{\sum_ip_i-1}}
  {p_1!\;...\;p_n!}(E_{\delta})^{p_1}(E_{2\delta})^{p_2}...
  (E_{n\delta})^{p_n}
\end{equation}
The order in the product (\ref{sl2R}) concides
with the above chosen normal order.

We now consider rank 3 case.
Fix the order in positive root system
$\Delta_+$ of $A_2^{(1)}$\,:
$\alpha$,\, $\alpha+\delta$,\, $\cdots$,\, $\alpha+m_1\delta$,\, $\cdots$,
\, $\alpha+\beta$,\,
$\alpha+\beta+\delta$,\, $\cdots$,\, $\alpha+\beta+m_2\delta$,\, $\cdots$,
\, $\beta$,\,
$\beta+\delta$,\, $\cdots$,\, $\beta+m_3\delta$,\, $\cdots$,\, $\delta$,\,
$2\delta$,\,...,\,
$k\delta$,\, $\cdots$,\, $\cdots$\, $(\delta-\beta)+l_1\delta$,\, $\cdots$,
\, $\delta-\beta$,\, $\cdots$,\,
$(\delta-\alpha)+l_2\delta$,\, $\cdots$,\, $\delta-\alpha$,\, $\cdots$,
\, $(\delta-\alpha-\beta)
+l_3\delta$,\, $\cdots$,\, $\delta-\alpha-\beta$,
where $m_i,k,l_i \geq 0\,,~~i=1,2,3$. Then one can show \cite{KT}\cite{ZG}
(see,
in particular, \cite{ZG}) that
the universal $R$-matrix for $U_q(A_2^{(1)})$
takes the explicit form
\begin{eqnarray}
R&=&\prod_{n\geq 0}~{\rm exp}_{q_\alpha}
\left ((q-q^{-1})(E_{\alpha+n\delta}
\otimes F_{\alpha+n\delta})\right )\nonumber\\
& &\cdot\prod_{n\geq 0}~{\rm exp}_{q_{\alpha+\beta}}
\left ((q-q^{-1})(E_{\alpha+\beta+n\delta}
\otimes F_{\alpha+\beta+n\delta})\right )\nonumber\\
& &\cdot \prod_{n\geq 0}~{\rm exp}_{q_\beta}
\left ((q-q^{-1})(E_{\beta+n\delta}
\otimes F_{\beta+n\delta})\right )\nonumber\\
& &\cdot {\rm exp}\left (\sum_{n>0}\sum^2_{i,j=1}
C^n_{ij}(q)(q-q^{-1})(E^{(i)}_{n\delta}\otimes F^{(j)}_{n\delta})
\right )\nonumber\\
& &\cdot \prod_{n\geq 0}~{\rm exp}_{q_{\beta}}
\left ((q-q^{-1})(E_{(\delta-\beta)+n\delta}
\otimes F_{(\delta-\beta)+n\delta})\right )\nonumber\\
& &\cdot \prod_{n\geq 0}~{\rm exp}_{q_\alpha}
\left ((q-q^{-1})(E_{(\delta-\alpha)+n\delta}
\otimes F_{(\delta-\alpha)+n\delta})\right )\nonumber\\
& &\cdot\prod_{n\geq 0}~{\rm exp}_{q_{\alpha+\beta}}
\left ((q-q^{-1})(E_{(\delta-\alpha-\beta)+n\delta}
\otimes F_{(\delta-\alpha-\beta)+n\delta})\right )\nonumber\\
& &\cdot q^{\sum^2_{i,j=1}\,(a^{-1}_{\rm sym})^{ij}h_i\otimes h_j
+c\otimes d+d\otimes c}\label{aR}
\end{eqnarray}
where $c=h_0+h_\psi$ with $\psi=\alpha+\beta$ being
the highest root of $A_2^{(1)}$ and
$(a_{\rm sym}^{ij})=\left (
\begin{array}{cc}
2 & -1\\
-1 & 2
\end{array} \right )$~;~
\begin{equation}
(C^n_{ij}(q))=(C^n_{ji}(q))=
 \frac{n}{[n]_q}\,\frac{[2]_q^2}{q^{2n}+1+q^{-2n}}\,\left (
\begin{array}{cc}
q^n+q^{-n} & (-1)^n\\
(-1)^n & q^n+q^{-n}
\end{array} \right )
\end{equation}
and  the Cartan-Weyl generators, $E_\gamma\,,~~F_\gamma=\theta(E_\gamma)
\,,~~\gamma\in\Delta_+$, are given by
($\alpha_i=\alpha,\,\beta,\,\alpha+\beta$ below) :
$E_{\alpha+\beta}=[E_\alpha\,,\,E_\beta]_q$ \,,~~
$E_{\delta-\alpha}=[E_\beta\,,\,E_{\delta-\alpha-\beta}]_q$ \,,~~
$E_{\delta-\beta}=[E_\alpha\,,\,E_{\delta-\alpha-\beta}]_q$ \,,~~
$\tilde{E}_\delta^{(i)}=[(\alpha_i,\alpha_i)]_q^{-1}[E_{\alpha_i},\,
  E_{\delta-\alpha_i}]_q$ \,,~~
$E_{\alpha_i+n\delta}=(-1)^n\left ({\rm ad}\tilde{E}_\delta^{(i)}\right )^n
  E_{\alpha_i}$ \,,~~
$E_{\delta-\alpha_i+n\delta}=\left ({\rm ad}\tilde{E}_\delta^{(i)}\right )^n
  E_{\delta-\alpha_i}$ \,,~~
$\tilde{E}_{n\delta}^{(i)}= [(\alpha_i,\alpha_i)]_q^{-1}[E_{\alpha_i
  +(n-1)\delta},\,E_{\delta-\alpha_i}]_q$ and
\begin{equation}
\tilde{E}^{(i)}_{n\delta}=\sum_{p_1+2p_2+\cdots+np_n=n}
  \frac{\left ( q^{(\alpha_i,\alpha_i)}-q^{-(\alpha_i,\alpha_i)}
  \right )^{\sum_ip_i-1}}
  {p_1!\;...\;p_n!}(E^{(i)}_{\delta})^{p_1}(E^{(i)}_{2\delta})^{p_2}...
  (E^{(i)}_{n\delta})^{p_n}
\end{equation}
The order in the product of (\ref{aR}) concides with the above defined order.
\vskip.1in
\noindent {\bf 3. Universal $R$-matrix with Spectral Parameter:}
We come to our main concerns in this Letter. We only list our results and
details will be published elsewhere \cite{ZG2}.

It can be shown that for any $z\in {\bf C}^\times$, there is a homomorphism
of algebras ${\rm ev}_z$: $U_q(A^{(1)}_1)\rightarrow U_q(A_1)$ given by:
${\rm ev}_z(E_\alpha)=E_\alpha\,,~~{\rm ev}_z(F_\alpha)=F_\alpha$\,,~~
${\rm ev}_z(h_\alpha)=h_\alpha\,,~~{\rm ev}_z(c)=0$\,,~~
${\rm ev}_z(E_{\delta-\alpha})=zF_\alpha$\,,~~${\rm ev}_z(F_{\delta-\alpha})
=z^{-1}E_\alpha$\,,~~
${\rm ev}_z(h_{\delta-\alpha})=-h_\alpha$.
Then from eq.(\ref{sl2R}) one obtains the
spectral-dependent universal $R$-matrix,
$R(x,y)\equiv ({\rm ev}_x\otimes {\rm ev}_y)R$, for $U_q(A_1)$~:
\begin{eqnarray}
R(x,y)&=&\prod_{n\geq 0}\,{\rm exp}_{q_\alpha}\left ( (q-q^{-1})\left (
  \frac{x}{y}\right )^n\left (q^{-nh_\alpha}E_\alpha\otimes F_\alpha
  q^{nh_\alpha}\right )\right )\nonumber\\
& &\cdot{\rm exp}\left ( \sum_{n>0}n[n]_{q_\alpha}^{-1}
  (q_\alpha-q_\alpha^{-1})\left (\frac{x}{y}\right )^n
  (E'_{n\delta}\otimes F'_{n\delta})\right )\nonumber\\
& &\cdot\prod_{n\geq 0}\;{\rm exp}_{q_\alpha}\left ((q-q^{-1})\left (
  \frac{x}{y}\right )^{n+1}\left (F_\alpha q^{-nh_\alpha}\otimes q^{nh_\alpha}
  E_\alpha\right )\right )\cdot
  q^{\frac{1}{2}h_\alpha\otimes h_\alpha}\label{loop-sl2R}
\end{eqnarray}
where $E'_{n\delta}$ and $F'_{n\delta}$ are determined by
the following equalities of formal power series:
\begin{eqnarray}
&&1+(q_\alpha-q_\alpha^{-1})\sum_{k=1}^\infty \tilde{E}'_
{k\delta}u^k={\rm exp}\left ( (q_\alpha-q_\alpha^{-1})\sum_{l=1}^\infty
E'_{l\delta}u^l\right )\nonumber\\
&&1-(q_\alpha-q_\alpha^{-1})\sum_{k=1}^\infty \tilde{F}'_
{k\delta}u^{-k}={\rm exp}\left ( -(q_\alpha-q_\alpha^{-1})\sum_{l=1}^\infty
F'_{l\delta}u^{-l}\right )\label{compare}
\end{eqnarray}
in which $\tilde{E}'_{n\delta}=[2]_q^{-1}(-1)^{n-1}q^{-(n-1)h_\alpha}\left (
 E_\alpha F_\alpha -q^{-2n}F_\alpha E_\alpha\right )$\,,~~
$\tilde{F}'_{n\delta}=\theta(\tilde{E}'_{n\delta})$.

We now apply (\ref{loop-sl2R}) to $V_{1/2}\otimes V_{1/2}$, where
$V_{1/2}$ is the fundamental representation of $U_q(A_1)$. One can
show that in this case
\begin{equation}
R_{1/2,1/2}(x,y)=f_q(x,y)\cdot \left (e_{11}+e_{44}+
  \frac{q^{-1}(y-x)}{y-q^{-2}x}(e_{22}+e_{33})+
  \frac{1-q^{-2}}{y-q^{-2}x}(ye_{23}+xe_{32})\right )\label{1/2,1/2}
\end{equation}
where (and below) $e_{ij}$ is the matrix satisfying $(e_{ij})_{kl}=
\delta_{ik}\delta_{jl}$ and
\begin{equation}
f_q(x,y)=q^{1/2}\cdot {\rm exp}\left (\sum_{n>0}\frac{q^n-q^{-n}}{q^n+q^{-n}}
\frac{(x/y)^n}{n}\right )
\end{equation}
We thus reproduce the well-known result \cite{Jimbo}, up to the scalar factor
$f_q(x,y)$. In \cite{KT} KT obtained (\ref{1/2,1/2}) directly from
(\ref{sl2R}).

We then turn to $U_q(A_2^{(1)})$. In this case,
one may show that for any $z\in {\bf C}^\times$, there is a homomorphism
of algebras ${\rm ev}_z$: $U_q(A_2^{(1)})\rightarrow U_q(A_2)$ given by:
${\rm ev}_z(E_\alpha)=E_\alpha\,,~~{\rm ev}_z(F_\alpha)=F_\alpha$\,,~~
${\rm ev}_z(h_\alpha)=h_\alpha$
\,,~~${\rm ev}_z(E_\beta)=E_\beta$\,,~~${\rm ev}_z(F_\beta)=F_\beta$\,,~~
${\rm ev}_z(h_\beta)=h_\beta$\,,~~
${\rm ev}_z(E_{\delta-\alpha-\beta})=zF_{\alpha+\beta}q^{(h_\beta-h_\alpha)/3}
$\,,~~${\rm ev}_z(F_{\delta-\alpha-\beta})=z^{-1}q^{(h_\alpha-h_\beta)/3}
  E_{\alpha+\beta}$\,,~~
${\rm ev}_z(h_{\delta-\alpha-\beta})=-h_{\alpha+\beta}$\,,~~${\rm ev}_z(c)=0$.
Thus, from  eq.(\ref{aR}) one is able to deduce the
explicitly spectral-dependent universal $R$-matrix for $U_q(A_2)$,
$R(x,y)\equiv ({\rm ev}_x\otimes {\rm ev}_y)R$, given by
\begin{eqnarray}
R(x,y)&=&\prod_{n\geq 0}~{\rm exp}_{q_\alpha}
\left ((q-q^{-1})\left (\frac{x}{y}\right )^n\left (q^{-nh_\alpha}E_\alpha
q^{-n(h_\alpha+2h_\beta)/3}\otimes q^{n(h_\alpha+2\beta)/3}F_\alpha
q^{nh_\alpha}\right )\right )\nonumber\\
& &\cdot\prod_{n\geq 0}~{\rm exp}_{q_{\alpha+\beta}}
\left ((q-q^{-1})\left (\frac{x}{y}\right )^n\left (q^{-nh_{\alpha+\beta}}
E_{\alpha+\beta}q^{n(h_\beta-h_\alpha)/3}
\otimes q^{n(h_\alpha-\beta)/3}F_{\alpha+\beta}q^{nh_{\alpha+\beta}}
\right )\right )\nonumber\\
& &\cdot \prod_{n\geq 0}~{\rm exp}_{q_\beta}
\left ((q-q^{-1})\left (\frac{x}{y}\right )^n\left (E'_{\beta+n\delta}
\otimes F'_{\beta+n\delta}\right )\right )\nonumber\\
& &\cdot {\rm exp}\left (\sum_{n>0}\sum^2_{i,j=1}
C^n_{ij}(q)(q-q^{-1})\left (\frac{x}{y}\right )^n
(E^{'(i)}_{n\delta}\otimes F^{'(j)}_{n\delta})\right )\nonumber\\
& &\cdot \prod_{n\geq 0}~{\rm exp}_{q_{\beta}}
\left ((q-q^{-1})\left (\frac{x}{y}\right )^{n+1}\left (E'_{(\delta-\beta)
+n\delta}\otimes F'_{(\delta-\beta)+n\delta}\right )\right )\nonumber\\
& &\cdot \prod_{n\geq 0}~{\rm exp}_{q_\alpha}
\left ((q-q^{-1})\left (\frac{x}{y}\right )^{n+1}\left (
q^{-(n+1)(h_\alpha+2h_\beta)/3}F_\alpha q^{-nh_\alpha}
\otimes q^{nh_\alpha}E_\alpha q^{(n+1)(h_\alpha+2h_\beta)/3}
\right )\right )\nonumber\\
& &\cdot\prod_{n\geq 0}~{\rm exp}_{q_{\alpha+\beta}}
\left ((q-q^{-1})\left (\frac{x}{y}\right )^{n+1}\left (q^{(n+1)(h_\beta-
h_\alpha)/3}F_{\alpha+\beta}q^{-nh_{\alpha+\beta}}\right .\right .\nonumber\\
& &\left .\left .~~~~~
\otimes q^{nh_{\alpha+\beta}}E_{\alpha+\beta}q^{(n+1)(h_\alpha-h_\beta)/3}
\right )\right )
\cdot q^{\sum^2_{i,j=1}\,(a^{-1}_{\rm sym})^{ij}h_i\otimes h_j}\label{loop-R}
\end{eqnarray}
where
\begin{eqnarray}
&&E'_{\beta+n\delta}=(-1)^n[2]_q^{-n}q^n\left \{\left ({\rm ad'}_{q^{-1}
  }{\cal E}\right )^nE_\beta\right \}q^{n(h_\beta-h_\alpha)/3}\nonumber\\
&&F'_{\beta+n\delta}=[2]_q^{-n}q^{n(h_\alpha-h_\beta)/3}
  \left ({\rm ad'}_{q^{-1}}{\cal F}\right )^nF_\beta\nonumber\\
&&E'_{(\delta-\beta)+n\delta}=[2]_q^{-n}q^{-n}\left \{\left ({\rm ad'}_q
  {\cal E}\right )^n({\rm ad'}_{q^{-2}}E_\alpha)F_{\alpha+\beta}
  \right \}q^{(n+1)(h_\beta-h_\alpha)/3}\nonumber\\
&&F'_{(\delta-\beta)+n\delta}=(-1)^n[2]_q^{-n}
  q^{(n+1)(h_\alpha-h_\beta)/3}
  \left ({\rm ad'}_q{\cal F}\right )^n({\rm ad'}_{q^2}
  E_{\alpha+\beta})F_\alpha
\end{eqnarray}
${\cal E}=({\rm ad'}_{q^{-1}}E_\beta)({\rm ad'}_{q^{-2}}E_\alpha)
  F_{\alpha+\beta}$~,~~${\cal F}=({\rm ad'}_q({\rm ad'}_{q^2}
  E_{\alpha+\beta})F_\alpha)F_\beta$ and
$({\rm ad'}_Q{\cal A})\cdot {\cal B}\equiv {\cal A}{\cal B}-Q{\cal B}{\cal A}$
{}~;~ $E^{'(i)}_{n\delta}$ and $F^{'(i)}_{n\delta}$ are determined by the
equalities of formal series: ($\alpha_i=\alpha,\,\beta$)
\begin{eqnarray}
&&1+(q_{\alpha_i}-q_{\alpha_i}^{-1})\sum_{k=1}^\infty\tilde{E}^{'(i)}_
  {k\delta}u^k={\rm exp}\left ((q_{\alpha_i}-q_{\alpha_i}^{-1})
  \sum_{l=1}^\infty E^{'(i)}_{l\delta}u^l\right )\nonumber\\
&&1-(q_{\alpha_i}-q_{\alpha_i}^{-1})\sum_{k=1}^\infty\tilde{F}^{'(i)}_
  {k\delta}u^{-k}={\rm exp}\left (-(q_{\alpha_i}-q_{\alpha_i}^{-1})
  \sum_{l=1}^\infty F^{'(i)}_{l\delta}u^{-l}\right )\label{primed}
\end{eqnarray}
in which
\begin{eqnarray}
&&\tilde{E}^{'(\alpha)}_{n\delta}=(-1)^{n-1}[2]^{-1}_q(E_\alpha F_\alpha
  -q^{-2n}F_\alpha E_\alpha)\,q^{-(n-1)h_\alpha}q^{-n(h_\alpha+2h_\beta)/3}
  \nonumber\\
&&\tilde{F}^{'(\alpha)}_{n\delta}=(-1)^{n-1}[2]^{-1}_q
  \,q^{(n-1)h_\alpha}q^{n(h_\alpha+2h_\beta)/3}(F_\alpha E_\alpha-q^{2n}
  E_\alpha F_\alpha)\nonumber\\
&&\tilde{E}^{'(\beta)}_{n\delta}=(-1)^n[2]_q^{-n}q^{n-2}\left \{
  \left ({\rm ad'}_{q^{-n+2}}{\cal F'}\right )\cdot\left ({\rm ad'}_{q^{-1}}
  {\cal E}\right )^{n-1}E_\beta\right \}q^{n(h_\beta-h_\alpha)/3}\nonumber\\
&&\tilde{F}^{'(\beta)}_{n\delta}=[2]_q^{-n}q^{n-1}q^{n(h_\alpha-h_\beta)/3}
  \left ({\rm ad'}_{q^{-n+2}}{\cal E'}\right )\cdot
  \left ({\rm ad'}_{q^{-1}}{\cal F}\right )^{n-1}F_\beta\nonumber\\
&&{\cal E'}=E_{\alpha+\beta}F_\alpha-q^2F_\alpha E_{\alpha+\beta}\,,~~~~
  {\cal F'}=E_{\alpha}F_{\alpha+\beta}-q^{-2}F_{\alpha+\beta} E_{\alpha}
\end{eqnarray}

We apply (\ref{loop-R}) to $V_{(3)}\otimes V_{(3)}$ in which
$V_{(3)}$ denotes the fundamental representation of $U_q(A_2)$.
One can show that (\ref{loop-R}) gives rise to
\begin{eqnarray}
R_{(3),(3)}(x,y)&=&f_q(x,y)\cdot \left (e_{11}+e_{55}+e_{99}
  +\frac{q^{-1}(y-x)}{y-q^{-2}x}(e_{22}+e_{33}+e_{44}+e_{66}+e_{77}+e_{88})
  +\right .\nonumber\\
& &\left .+\frac{y(1-q^{-2})}{y-q^{-2}x}(e_{24}+e_{37}+e_{68})
  +\frac{x(1-q^{-2})}{y-q^{-2}x}(e_{42}+e_{73}+e_{86})\right )\label{33}
\end{eqnarray}
where
\begin{equation}
f_q(x,y)=q^{2/3}\cdot {\rm exp}\left (\sum_{n>0}\frac{q^{2n}-q^{-2n}}{q^{2n}+1
 +q^{-2n}}\,\frac{(x/y)^n}{n}\right )
\end{equation}
Eq.(\ref{33}) is nothing but the well-known result \cite{Jimbo},
up to the scalar factor $f_q(x,y)$.

We then apply (\ref{loop-R}) to a very interesting case: $V_{(8)}\otimes
V_{(8)}$, where $V_{(8)}$ stands for the adjoint representation of
$U_q(A_2)$. This is simplest nontrivial
example where the tensor product decomposition is with finite multiplicity.
It can be shown that spectral-dependent $R$-matrix for the adjoint
representation takes the following compact and explicit form,
\begin{eqnarray}
R_{(8),(8)}(x,y)&=&\left \{1+(q-q^{-1})\sum_{n=0}^\infty\left (\frac{x}{y}
  \right )^n \left (E'_{\alpha+n\delta}\otimes F'_{\alpha+n\delta}\right )
  +y^2f(e_{36}\otimes e_{63})\right \}\nonumber\\
& &\cdot\left \{1+(q-q^{-1})\sum_{n=0}^\infty\left (\frac{x}{y}\right )^n
  \left (E'_{\alpha+\beta+n\delta}\otimes F'_{\alpha+\beta+n\delta}\right )
  +y^2f(e_{18}\otimes e_{81})\right \}\nonumber\\
& &\cdot\left \{1+(q-q^{-1})\sum_{n=0}^\infty\left (\frac{x}{y}\right )^n
  \left (E'_{\beta+n\delta}\otimes F'_{\beta+n\delta}\right )
  +y^2f'(e_{27}\otimes e_{72})\right \}\nonumber\\
& &\cdot\left \{{\rm imaginary~root~vectors~contribution}\right \}\nonumber\\
& &\cdot\left \{1+(q-q^{-1})\sum_{n=0}^\infty\left (\frac{x}{y}\right )^{n+1}
  \left (E'_{(\delta-\beta)+n\delta}\otimes F'_{(\delta-\beta)+n\delta}\right )
  +x^2g'(e_{72}\otimes e_{27})\right \}\nonumber\\
& &\cdot\left \{1+(q-q^{-1})\sum_{n=0}^\infty\left (\frac{x}{y}\right )^{n+1}
  \left (E'_{(\delta-\alpha)+n\delta}\otimes F'_{(\delta-\alpha)+n\delta}
  \right )+x^2g(e_{63}\otimes e_{36})\right \}\nonumber\\
& &\cdot\left \{1+(q-q^{-1})\sum_{n=0}^\infty\left (\frac{x}{y}\right )^{n+1}
  \left (E'_{(\delta-\alpha-\beta)+n\delta}\otimes
  F'_{(\delta-\alpha-\beta)+n\delta}\right )
  +x^2g(e_{81}\otimes e_{18})\right \}\nonumber\\
& &\cdot\left \{q^2\sum_{i=1}^8\left (1-\delta_{i4}-\delta_{i5})
  (e_{ii}\otimes e_{ii}\right )
  +q\left (e_{11}\otimes e_{22}+e_{11}\otimes e_{33}+e_{22}\otimes e_{66}+
  \right .\right .\nonumber\\
& &+\left .e_{33}\otimes e_{77}+e_{66}\otimes e_{88}+e_{77}\otimes e_{88}+
  \{\leftrightarrow\}\right )
  +q^{-1}\left (e_{11}\otimes e_{66}+e_{11}\otimes e_{77}+\right .\nonumber\\
& &+e_{22}\otimes e_{33}+
  e_{22}\otimes e_{88}+e_{33}\otimes e_{88}+e_{66}\otimes e_{77}+
  \{\leftrightarrow\})+\nonumber\\
& &\left .+  q^{-2}\left (e_{11}\otimes e_{88}+e_{22}\otimes e_{77}+
  e_{33}\otimes e_{66}+\{\leftrightarrow\}\right )\right \}\label{88R}
\end{eqnarray}
where $f=[2]_qq^{-1}(q-q^{-1})^2\,(y+q^4x)/\{(y^2-x^2)(y-q^2x)\}$,~
$g=[2]_qq^{-1}(q-q^{-1})^2\,(y+q^{-2}x)/\{(y^2-x^2)(y-q^{-2}x)\}$,~
$f'=[2]_q^{-2}(q-q^{-1})^2(y^2-x^2)^{-1}(q^{-1}[2]_q^3+F(q;x,y))$,~
$g'=[2]_q^{-2}(q-q^{-1})^2(y^2-x^2)^{-1}(q^{-1}[2]_q^3+F(q^{-1};x,y))$,~
$F(q,x,y)=q^{-4}x/(y-q^{-4}x)+2[3]_qx/(y-x)+[3]_q^2q^4x/(y-q^4x)$,~
and $"\{\leftrightarrow\}"$ denotes the interchange to $X\otimes Y$
of the quantities within the same bracket;
\begin{eqnarray}
&&E'_{\alpha+n\delta}=(-1)^n\left \{q^{-2n}e_{12}+[2]_q^{1/2}q^{-2n}e_{34}+
  [2]_q^{1/2}e_{46}+e_{78}\right \}\nonumber\\
&&F'_{\alpha+n\delta}=(-1)^n\left \{q^{2n}e_{21}+[2]_q^{1/2}q^{2n}e_{43}+
  [2]_q^{1/2}e_{64}+e_{87}\right \}\nonumber\\
&&E'_{\alpha+\beta+n\delta}=(-1)^n\left \{-[2]_q^{-1/2}q^{-2n-2}e_{14}+
  \left ([3]_q/[2]_q\right )^{1/2}q^{-2n}e_{15}-q^{-1}e_{26}
  +\right .\nonumber\\
&&~~~~~~~~~~~~~~~+\left .q^{-2n}e_{37}+[2]_q^{-1/2}qe_{48}-\left (
  [3]_q/[2]_q\right )^{1/2}q^{-1}e_{58}\right \}\nonumber\\
&&F'_{\alpha+\beta+n\delta}=(-1)^n\left \{-[2]_q^{-1/2}q^{2n+2}e_{41}+
  \left ([3]_q/[2]_q\right )^{1/2}q^{2n}e_{51}-qe_{62}
  +\right .\nonumber\\
&&~~~~~~~~~~~~~~~+\left .q^{2n}e_{73}+[2]_q^{-1/2}q^{-1}e_{84}-\left (
  [3]_q/[2]_q\right )^{1/2}qe_{85}\right \}\nonumber\\
&&E'_{\beta+n\delta}=q^ne_{13}+[2]_q^{-1/2}q^ne_{24}+
  \left ([3]_q/[2]_q\right )^{1/2}q^{-3n}e_{25}+\nonumber\\
&&~~~~~~~~~~~+[2]_q^{-1/2}q^{-3n}e_{47}+\left ([3]_q/[2]_q\right )^{1/2}
  q^ne_{57}+q^{-3n}e_{68}\nonumber\\
&&F'_{\beta+n\delta}=q^{-n}e_{31}+[2]_q^{-1/2}q^{-n}e_{42}+
  \left ([3]_q/[2]_q\right )^{1/2}q^{3n}e_{52}+\nonumber\\
&&~~~~~~~~~~~+[2]_q^{-1/2}q^{3n}e_{74}+\left ([3]_q/[2]_q\right )^{1/2}
  q^{-n}e_{75}+q^{3n}e_{86}\nonumber\\
&&E'_{(\delta-\beta)+n\delta}=-q^{n+2}e_{31}-[2]_q^{-1/2}q^{n+3}e_{42}-
  \left ([3]_q/[2]_q\right )^{1/2}q^{-3n-1}e_{52}\nonumber\\
&&~~~~~~~~~~~~~~~-[2]_q^{-1/2}q^{-3n-3}e_{74}-\left (
  [3]_q/[2]_q\right )^{1/2} q^{n+1}e_{75}-q^{-3n-2}e_{86}\nonumber\\
&&F'_{(\delta-\beta)+n\delta}=-q^{-n-2}e_{13}-[2]_q^{-1/2}q^{-n-3}e_{24}-
  \left ([3]_q/[2]_q\right )^{1/2}q^{3n+1}e_{25}\nonumber\\
&&~~~~~~~~~~~~~~~~-[2]_q^{-1/2}q^{3n+3}e_{47}-\left (
  [3]_q/[2]_q\right )^{1/2} q^{-n-1}e_{57}-q^{3n+2}e_{68}\nonumber\\
&&E'_{(\delta-\alpha)+n\delta}=(-1)^n\left \{q^{-2n-1}e_{21}+
  [2]_q^{1/2}q^{-2n}e_{43}+
  [2]_q^{1/2}e_{64}+qe_{87}\right \}\nonumber\\
&&F'_{(\delta-\alpha)+n\delta}=(-1)^n\left \{q^{2n+1}e_{12}+
  [2]_q^{1/2}q^{2n}e_{34}+
  [2]_q^{1/2}e_{46}+q^{-1}e_{78}\right \}\nonumber\\
&&E'_{(\delta-\alpha-\beta)+n\delta}=(-1)^n\left \{-[2]_q^{-1/2}q^{-2n+2}e_{41}
  +\left ([3]_q/[2]_q\right )^{1/2}q^{-2n}e_{51}-q^2e_{62}\right .\nonumber\\
&&~~~~~~~~~~~~~~~+\left .q^{-2n-1}e_{73}+[2]_q^{-1/2}q^{-1}e_{84}-\left (
  [3]_q/[2]_q\right )^{1/2}qe_{85}\right \}\nonumber\\
&&F'_{(\delta-\alpha-\beta)+n\delta}=(-1)^n\left \{-[2]_q^{-1/2}q^{2n-2}e_{14}+
  \left ([3]_q/[2]_q\right )^{1/2}q^{2n}e_{15}-q^{-2}e_{26}
  +\right .\nonumber\\
&&~~~~~~~~~~~~~~~+\left .q^{2n+1}e_{37}+[2]_q^{-1/2}qe_{48}-\left (
  [3]_q/[2]_q\right )^{1/2}q^{-1}e_{58}\right \}\nonumber\\
&&\{{\rm imaginary~root~vectors~contribution}\}=a'/a\,\sum_{i=1}^8\left (
  1+(b/b'-1)\delta_{i4}+(c/c'-1)\delta_{i5}\right )(e_{ii}\otimes e_{ii})+
  \nonumber\\
&&~~~~~~~~~~~~~~~~+  a'(e_{11}\otimes e_{22}+e_{11}\otimes e_{33})
  +aa'(e_{11}\otimes e_{44})
  +a'c(e_{11}\otimes e_{55})+a(e_{11}\otimes e_{66})+\nonumber\\
&&~~~~~~~~~~~~~~~~+aa'c(e_{11}\otimes
  e_{77})+ac(e_{11}\otimes e_{88})+1/a(e_{22}\otimes e_{11})+
  1/b'(e_{22}\otimes e_{33})+\nonumber\\
&&~~~~~~~~~~~~~~~~+  a'/b'(e_{22}\otimes e_{44})+a'c(e_{22}\otimes
  e_{55})+a'(e_{22}\otimes e_{66})+aa'c/b'(e_{22}\otimes e_{77})+\nonumber\\
&&~~~~~~~~~~~~~~~~+aa'c(e_{22}\otimes e_{88})+1/a(e_{33}\otimes e_{11})
  +b(e_{33}\otimes e_{22})
  +a'b(e_{33}\otimes e_{44})+\nonumber\\
&&~~~~~~~~~~~~~~~~+  ab(e_{33}\otimes e_{66})+a'(e_{33}\otimes e_{77})+
  a(e_{33}\otimes e_{88})+1/(aa')(e_{44}\otimes e_{11})+\nonumber\\
&&~~~~~~~~~~~~~~~~+  b/a(e_{44}\otimes
  e_{22})+1/(ab')(e_{44}\otimes e_{33})+a'b(e_{44}\otimes e_{66})+a'/b'
  (e_{44}\otimes e_{77})+\nonumber\\
&&~~~~~~~~~~~~~~~~+aa'(e_{44}\otimes e_{88})+1/(ac')(e_{55}\otimes e_{11}+
  e_{55}\otimes e_{22})+a'c(e_{55}\otimes e_{77}+e_{55}\otimes e_{88})
  +\nonumber\\
&&~~~~~~~~~~~~~~~~+  1/a'(e_{66}\otimes e_{11})+
  1/a(e_{66}\otimes e_{22})+1/(a'b')(e_{66}\otimes e_{33})+1/(ab')(e_{66}
  \otimes e_{44})+\nonumber\\
&&~~~~~~~~~~~~~~~~+  1/b'(e_{66}\otimes e_{77})+a'(e_{66}\otimes e_{88})+
  1/(aa'c')(e_{77}\otimes e_{11})+
  b/(aa'c')(e_{77}\otimes e_{22})+\nonumber\\
&&~~~~~~~~~~~~~~~~+  1/a(e_{77}\otimes e_{33})+b/a(e_{77}
  \otimes e_{44})+1/(ac')(e_{77}\otimes e_{55})+b(e_{77}\otimes e_{66})+
  \nonumber\\
&&~~~~~~~~~~~~~~~~+  a'(e_{77}\otimes e_{88})+
  1/(a'c')(e_{88}\otimes e_{11})+1/(aa'c')(e_{88}\otimes e_{22})+
  1/a'(e_{88}\otimes e_{33})+\nonumber\\
&&~~~~~~~~~~~~~~~~+  1/(aa')(e_{88}\otimes e_{44})+1/(ac')(e_{88}\otimes
  e_{55})+1/a(e_{88}\otimes e_{66}+e_{88}\otimes e_{77})\label{generators}
\end{eqnarray}
in which we have defined
$a=(y-q^2x)/(y-x)$\,,~~$a'=(y-q^{-2}x)/(y-x)$\,,~~$b=(y-q^4x)/(y-q^2x)$\,,~~
$b'=(y-q^{-4}x)/(y-q^{-2}x)$\,,~~$c=(y-q^6x)/(y-q^4x)$\,,~~
$c'=(y-q^{-6}x)/(y-q^{-4}x)$.
We remark that (\ref{88R}) is an extreamly explicit formula: the sums in
(\ref{88R}) can be easily worked out. For compactness, we leave them as
the present form.
\vskip.1in
\noindent {\bf 4. Concluding Remarks:}
In this letter, we have obtained the spectral-dependent universal
$R$-matrix for $U_q(A_1)$
and $U_q(A_2)$.  As their
applications, we reproduce from them some well-known results and obtain
an extreamly explicit and compact formula for spectral-dependent $R$-matrix
in the adjoint representation of $U_q(A_2)$.
We believe that our results will be useful in: (i)
one dimensional open spin chains \cite{Davies et al}, (ii)
quantization of the conformal affine Liouville (and Toda) theories
\cite{Bonora} and (iii) $q$-deformed WZNW CFT's \cite{Frenkel}.

\vskip.3in
One of the authors (Y.Z.Z.) is grateful to Anthony John Bracken for contineous
interest, suggestions and discussions, to V.N.Tolstoy for communication of
his papers on quantum groups and to R.Cuerno for email correspondences.
After completing this paper,  Zhong-Qi Ma informed Y.Z.Z.  of
an earlier paper by him and his coworkers \cite{Ma et al}
in which part of the results in the present paper --
the quantum $R$-matrix  associated with the
adjoint reps of $U_q(A_2)$ -- was also obtained using the conventional method
in the literature -- directly solving Jimbo-type equations with the help
of quantum CG coefficients (projection operators) of $U_q(A_2)$.
The present paper, however, presents a new and different procedure which
avoids employing the knowledge of the quantum CG coefficients which are
not generally available. Y.Z.Z. thanks Zhong-Qi Ma for e-mail correspondences
and for relevant comments. The financial support from Australian Research
Council is gratefully acknowledged.
\vskip.3in

\end{document}